\documentclass[review]{elsarticle}

\usepackage{hyperref}
\usepackage{amsmath,amssymb,amsfonts}
\usepackage{algorithmic}
\usepackage{ntheorem}
\usepackage{caption}
\usepackage{graphicx}
\usepackage{textcomp}
\usepackage{indentfirst}
\usepackage{float}
\usepackage{bm}
\usepackage{tagging}
\usepackage{amsfonts,amssymb}
\usepackage{amsmath}
\usepackage{tablists}
\usepackage{subfigure}
\usepackage{stfloats}
\usepackage{color}
\usepackage{appendix}
\DeclareMathAlphabet\mathbfcal{OMS}{cmsy}{b}{n}

\journal{Journal of \LaTeX\ Templates}

\begin{document}
\captionsetup[figure]{labelfont={bf},labelformat={default},labelsep=period,name={Fig.}}

\begin{frontmatter}

\title{Optimal Fractional Fourier Filtering in Time-vertex Graphs signal processing \tnoteref{mytitlenote} }
\tnotetext[mytitlenote]{This work has been supported by National Natural Science Foundations of China (No.62071242)}

\author{Zirui~Ge, Haiyan~Guo, Tingting~Wang, Zhen~Yang\corref{mycorrespondingauthor} }
\address{School of Communication and Information Engineering, Nanjing University of Posts and Telecommunications, Nanjing 2100023, China}
\fntext[myfootnote]{E-mail address: yangz@njupt.edu.cn (Z. Yang)}
%%1019010430@njupt.edu.cn (Z. Ge), guohy@njupt.edu.cn (H.Guo), 2018010215@njupt.edu.cn (T. Wang), 

%%ead{ }
\cortext[mycorrespondingauthor]{Corresponding author}
%\ead{ This work has been supported by National Natural Science Foundations of China (No.61671252, No.61271335, No.61901229), the Natural Science Research of}


\begin{abstract}
Graph signal processing (GSP) is an effective tool in dealing with data residing on irregular domains. In GSP, the optimal graph filter is one of the essential techniques, owing to its ability to recover the original signal from the distorted and noisy version. However, most current research focuses on static graph signals and ordinary space/time or frequency domains. The time-varying graph signals have a strong ability to capture the features of real-world data, and fractional domains can provide a more suitable space to separate the signal and noise. In this paper, the optimal time-vertex graph filter and its Wiener-Hopf equation are developed, using the product graph framework. Furthermore, the optimal time-vertex graph filter in fractional domains is also developed, using the graph fractional Laplacian operator and graph fractional Fourier transform. Numerical simulations on real-world datasets will demonstrate the superiority of the optimal time-vertex graph filter in fractional domains over the optimal time-vertex graph filter in ordinary domains and the optimal static graph filter in fractional domains.
\end{abstract}

\end{frontmatter}

\section{Introduction}
\par Graph signal processing extends classical digital signal processing (DSP) technologies to signals that reside on irregular structures \cite{Ref1,Ref2}. Some important concepts in DSP have been extended to GSP, such as the graph shift, graph Fourier transform (GFT) \cite{Ref3}, graph filter \cite{Ref4,Ref5,Ref6}, graph convolution and modulation \cite{Ref9}, etc. Among these concepts, Graph filter and GFT are important tools to process the graph signal. They have shown their advantage in graph neural networks \cite{Ref9,Ref10}, graph signal denoising \cite{Ref11,Ref12,Ref13}, graph signal recovery \cite{Ref14}, speech enhancement \cite{Ref15,Ref16} and others.

\par Similar to the frequency of digital signal processing (DSP), the definition of a Fourier transform for graph signals is introduced by the graph Fourier transform (GFT), and the GFT has been defined in two approaches. The first way is to use the adjacency matrix as the graph shift operator (GSO), and the obtained graph spectrum is rooted in the algebraic signal processing theory \cite{Ref26,Ref27}. The other approach for GFT is to use the graph Laplacian matrix as the graph shift operator. This technique is based on the spectral graph theory \cite{Ref28,Ref29,Ref30,Ref31}. 

\par The GFT has been extended to graph fractional Fourier transform (GFRFT) \cite{Ref43, Ref44}. The ath order GFRFT is the $ath$ power of the ordinary GFT. The GFRFT becomes an identity operator for $a=0$, and the GFRFT becomes the ordinary GFT for $a=1$. The GFRFT provides an intermediate domain between the original graph signal and GFT. The $ath$ order graph fractional shift operator (GFSO) based on the Laplacian matrix is composed of the $ath$ order GFRFT basis and the $ath$ power of the original eigenvalue matrix. Similarly, GFSO becomes an identity matrix with $a=0$, and ordinary graph shift operator with $a=1$ and GFSO also satisfies linear shift-invariant property [32,33]. The GFRFT and GFSO provide an extension of the ordinary GFT and GSO. With the additional degree of freedom, we can use the GFRFT and GFSO to improve the performance of the ordinary GFT and GSO in some fields, such as the optimal graph filter design.

\par The optimal graph filter as an essential part of graph filters has been discussed in \cite{Ref25}. The authors in \cite{Ref25} introduced the Wiener-Hopf equation for graph signals and the least-square graph filter based on the linear shift-invariance. The authors in \cite{Ref41} extended the optimal graph filter to fractional Fourier domains and introduced the optimal fractional Fourier domain filtering problem to GSP. 

\par In \cite{Ref34, Ref35}, the authors have extended the static graph signal processing framework to the time-vertex graph signal framework based on the Cartesian product graph model. In this framework, joint Fourier transform (JFT) is defined by GFT and classical discrete Fourier transform (DFT). Based on JFT, the joint graph filter (JGF) is defined in the joint frequency domain, i.e., graph frequency and angular frequency, and modeled as a polynomial of the Cartesian product of the Laplacian matrix of the underlying graph and the cycle graph \cite{Ref35, Ref36}. With the advantage of simultaneously capturing variations over graph topology and time, JGF can obtain better performance in many fields, including denoising in \cite{Ref37, Ref23}, reconstruction of time-vertex signals in \cite{Ref40}, and others.

\par However, the time-vertex graph signal processing framework has not been applied to the optimal fractional Fourier graph filtering model in \cite{Ref41}, which may perform better than the original model. Similarly, the graph Wiener-Hopf equation in \cite{Ref25} does not possess a time-vertex version. These gaps in the research on optimal graph filtering motive us to investigate the optimal fractional Fourier time-vertex (OFrFTv) graph filter.

\par In this paper, we first formulate the optimal time-vertex graph filter and obtain the optimal graph filter coefficients. The graph Wiener-Hopf equation in the time-vertex version is obtained in this processing. Based on the GFSO and the optimal time-vertex graph filter, we introduce the OFrFTv graph filter. 

\par The main contributions of this paper are summarized as follows:

\par 1)	We first extend the optimal graph filter on the static graph to the time-vertex graph. Then we obtain a closed-form solution to the optimal filtering problem on the time-vertex graph that is the Wiener-Hopf equation in the time-vertex version.

\par 2) Based on the GFSO, we further extend the ordinary optimal time-vertex graph filter to the optimal time-vertex graph filter in fractional domains. Specifically, we use fractional Laplacian operators based on the graph of underlying data and the cycle graph to build an OFrFTv graph filter.

\par 3) We present applications of the proposed graph filter in denoising. The comparison technique is the optimal fractional Fourier static (OFrFS) graph filter \cite{Ref41}. We take the Tikhonov regularization \cite{Ref35} and recursive graph median filter \cite{Ref42} as our preprocessing tools. The final results show that our graph filter outperforms the OFrFS graph filter and improves preprocessed results, especially in a low SNR.

\par The outline of the paper is as follows. Section II introduces the basic concepts of signal processing, fractional Fourier, and fractional shift operator for graphs. Section III presents the optimal graph filter on the time-vertex graph. In Section IV, we discuss the optimal fractional Fourier time-vertex graph filter. Section V discusses the implementation of the proposed graph filter. Our simulation results are provided in Section VI, while Section VII concludes the paper.

\par \emph{Notations}: The lowercase boldface letters $\boldsymbol{a}$ indicate vectors, while the uppercase boldface letters $\boldsymbol{A}$ denote matrices. Given a vector $\boldsymbol{a}$ or matrix $\boldsymbol{A}$, $\left[ \boldsymbol{a} \right] _i$ or $\left[ \boldsymbol{A} \right] _{i,j}$ denotes the $ith$ entry of $\boldsymbol{a}$ or $(i,j)th$ entry of $\boldsymbol{A}$. We indicate the 2-norm of the vector $\boldsymbol{a}$ and matrix $\boldsymbol{A}$ by $\left\| \boldsymbol{a} \right\| ^2$ and $\left\| \boldsymbol{A} \right\| _{2}^{2}$, respectively. We denote by $\boldsymbol{A}^{\left( a \right)}$ the $a\mathrm{th}$ order fractional Laplacian matrix of $\boldsymbol{A}$. $\otimes $ represents the Kronecker product. $\boldsymbol{A}^H$, $\overline{\boldsymbol{A}}$ and $\boldsymbol{A}^T$ represent the Hermitian, conjugation and transpose of $\boldsymbol{A}$.

\section{Related Work}
\subsection{Temporal Graph Signals.} Consider an undirected graph $\mathcal{G}=\left( \bm{\mathcal{V}},\bm{\mathcal{E}} \right) $ with $N$ nodes and $M$ edges, where $\bm{\mathcal{V}}=\left[	\upsilon_1,\upsilon_2,\cdots,\upsilon_N \right] $ denotes the set of $N$ vertices and $\bm{\mathcal{E}}$ denotes the set of $M$ edges. The underlying structure of $\mathcal{G}$ is described by the adjacency matrix $\boldsymbol{A}_{\mathcal{G}}\in R^{N\times N}$ or by the Laplacian matrix $\boldsymbol{L}_{\mathcal{G}}=\boldsymbol{D}_{\mathcal{G}}-\boldsymbol{A}_{\mathcal{G}}$, where $\boldsymbol{D}_{\mathcal{G}}$ is the degree diagonal matrix. Since $\mathcal{G}$ is undirected, the edge between $\upsilon_i$ and $\upsilon_j$ is same as that between $\upsilon_j$ and $\upsilon_i$. It is worth noting that the Laplacian matrix $\boldsymbol{L}_{\mathcal{G}}$ is symmetric, and its eigen-decomposition can be further expressed as $\boldsymbol{L}_{\mathcal{G}}=\boldsymbol{U}_{\mathcal{G}}\boldsymbol{\Lambda }_{\mathcal{G}}\boldsymbol{U}_{\mathcal{G}}^{H}$, where $\boldsymbol{U}_{\mathcal{G}}$ is the eigenvector matrix and $\boldsymbol{\Lambda }_{\mathcal{G}}$ is the diagonal matrix.
\par Suppose that the signals on a graph sampled at $T$ successive unit lengths are presented by a $\boldsymbol{X}\in \mathbb{R} ^{N\times T}$, with $\boldsymbol{X}_{nt}$ being the value of node $\upsilon _n$ at the time instant $t$.The JFT is written as
$$
\mathrm{JFT} \left\{ \boldsymbol{X} \right\} =\boldsymbol{U}_{\mathcal{G}}\boldsymbol{XU}_{\mathcal{T}}^{H},\eqno(1)
$$

\par \noindent where $\boldsymbol{U}_{\mathcal{G}}$ is the GFT matrix, $\boldsymbol{U}_{\mathcal{T}}$ is the DFT matrix. Express (1) in the vector form, and the transform becomes
$$
\boldsymbol{\hat{x}}=\mathrm{JFT}\left\{ \boldsymbol{x} \right\} =\left( \boldsymbol{U}_{\mathcal{T}}\otimes \boldsymbol{U}_{\mathcal{G}} \right) \boldsymbol{x}=\boldsymbol{U}_J\boldsymbol{x},\eqno(2)
$$
\noindent where $\boldsymbol{x}=\mathrm{vec}\left( \boldsymbol{X} \right) \in \mathbb{R} ^{NT}$, and $\mathrm{vec}\left( \cdot \right) $ is the vectorizing operator. In the last equation, we set $\boldsymbol{U}_J=\boldsymbol{U}_{\mathcal{T}}\otimes \boldsymbol{U}_{\mathcal{G}}$. The inverse joint time-vertex Fourier transform (IJFT) is $\boldsymbol{U}_{J}^{-1}=\boldsymbol{U}_{\mathcal{T}}^{-1}\otimes \boldsymbol{U}_{\mathcal{G}}^{-1}$. In this paper, we focus on undirected graph, and $\boldsymbol{U}_{\mathcal{G}}$ is a unitary matrix and $\boldsymbol{U}_{J}^{H}=\boldsymbol{U}_{J}^{-1}$.
\par The joint filter is
$$
\boldsymbol{H}=h\left( \begin{array}{c}
	\boldsymbol{L}_{\mathcal{T}},\boldsymbol{L}_{\mathcal{G}}\\
\end{array} \right) =\sum_{p=0,q=0}^{P-1,Q-1}{c_{p,q}\boldsymbol{L}_{\mathcal{T}}^{p}\otimes \boldsymbol{L}_{\mathcal{G}}^{q}},\eqno(3)
$$
\noindent and its corresponding frequency response is presented as 
$$
h\left( \boldsymbol{\varLambda }_{\mathcal{T}},\boldsymbol{\varLambda }_{\mathcal{G}} \right) =\sum_{p=0,q=0}^{P-1,Q-1}{c_{p,q}\boldsymbol{\varLambda }_{\mathcal{T}}^{p}\otimes \boldsymbol{\varLambda }_{\mathcal{G}}^{q}},\eqno(4)
$$
\noindent where $\boldsymbol{L}_{\mathcal{T}}=\boldsymbol{I}_{\mathcal{T}}-\boldsymbol{A}_{\mathcal{T}}$, and $\boldsymbol{A}_{\mathcal{T}}$ is the adjacency matrix of the cycle graph $\mathcal{T} $. More details can be found in [35].
\subsection{Graph Fractional Fourier Transform (GFRFT).} The ath GFRFT matrix of a GFT matrix $\boldsymbol{U}_{\mathcal{G}}^{H}$ is defined as

$$
\boldsymbol{F}^{\left( a \right)}=\boldsymbol{PJ}^a\mathbf{P}^{-1}, \eqno(5)
$$
\par \noindent where $0\leqslant a\leqslant 1$, and discrete fractional Fourier transform keeps unitarity, i.e., $\boldsymbol{F}^{\left( a \right)}$ is unitary. $\boldsymbol{P}$ is the Jordan eigenvector matrix of $\boldsymbol{U}_{\mathcal{G}}^{H}$, and
$$
\boldsymbol{U}^H=\boldsymbol{PJP}^{-1}.
$$
\par For a graph signal $\boldsymbol{x}\in \mathbb{C} ^N$, its $ath$ GFRFT is denoted as
$$
\boldsymbol{\hat{x}}=\boldsymbol{F}^{\left( a \right)}\boldsymbol{x},\eqno(6)
$$
and its inverse GFRFT is
$$
\boldsymbol{x}=\left( \boldsymbol{F}^{\left( a \right)} \right) ^{-1}\boldsymbol{\hat{x}}=\boldsymbol{F}^{\left( -a \right)}\boldsymbol{\hat{x}}=\boldsymbol{F}^{\left( a \right) H}\boldsymbol{\hat{x}}.
$$
When $a=0$ and $a=1$, the GFRFT reduces to identity matrix $\boldsymbol{I}$ and GFT matrix $\boldsymbol{U}_{\mathcal{G}}^{H}$. More details about GFRFT can be found in \cite{Ref43, Ref44}.
\par \noindent \textbf{C. The Graph Fractional Shift Operator (GFSO).} The definition of GFSO is as follows
$$
\boldsymbol{L}^{\left( a \right)}=\boldsymbol{F}^{\left( a \right)}{\boldsymbol{RF}^{\left( a \right)}}^H, \eqno(7)
$$
where $\boldsymbol{R}=\boldsymbol{\Lambda }^a$ and $\boldsymbol{\Lambda }$ is the eigenvalue matrix of the Laplacian matrix $\boldsymbol{L}_{\mathcal{G}}$, and more details about GFSO can be found in \cite{Ref32,Ref33}.

\section{Optimal Graph Filter on Temporal Graph Signals}
This section introduces the joint optimal graph filter on temporal graph signals. We aim to obtain the coefficients of the proposed optimal filter. The Wiener-Hopf equation of the time-vertex graph version in this solving processing is obtained.
\par Assume that $\boldsymbol{X}=\left[ \boldsymbol{x}_1,\cdots ,\boldsymbol{x}_T \right] \in \mathbb{R} ^{N\times T}$ is the temporal graph signal and $\boldsymbol{Y}=\left[ \boldsymbol{y}_1,\cdots ,\boldsymbol{y}_T \right] \in \mathbb{R} ^{N\times T}$ is a noisy measurement of graph signal $\boldsymbol{X}$, where $\boldsymbol{y}_t=\boldsymbol{x}_t+\boldsymbol{n}_t$, and $\boldsymbol{n}_t=\left[n_{1t},n_{2t},	\cdots, n_{Nt}\right] ^T$ is i.i.d. zeros mean white Gaussian noise. We aim to design a time-vertex graph filter to obtain an estimated graph signal $\boldsymbol{\tilde{x}}=\boldsymbol{Hy}$ that has minimum mean-squared error (MSE) $\mathbb{E} \left\{ \left\| \boldsymbol{\tilde{x}}_t-\boldsymbol{x}_t \right\| _{2}^{2} \right\} $. Then, the optimization problem can be written as
$$
\min_{\boldsymbol{H}} \mathbb{E} \left\{ \left\| \boldsymbol{Hy}-\boldsymbol{x} \right\| _{2}^{2} \right\} , \eqno(8)
$$
where $\boldsymbol{H}=h\left( \boldsymbol{L}_{\mathcal{T}},\boldsymbol{L}_{\mathcal{G}} \right) =\sum_{p=0,q=0}^{P-1,Q-1}{c_{p,q}\boldsymbol{L}_{\mathcal{T}}^{p}\otimes \boldsymbol{L}_{\mathcal{G}}^{q}}
$, $\boldsymbol{y}=\mathrm{vec}\left\{ \boldsymbol{Y} \right\} $, and $\boldsymbol{x}=\mathrm{vec}\left\{ \boldsymbol{X} \right\} $.
\par Let
$$
\begin{gathered}
\boldsymbol{B}=\left[\boldsymbol{L}_{\mathcal{T}}^{0} \otimes \boldsymbol{L}_{\mathcal{G}}^{0} \boldsymbol{y}, \boldsymbol{L}_{\mathcal{T}}^{0} \otimes \boldsymbol{L}_{\mathcal{G}}^{1} \boldsymbol{y}, \cdots, \boldsymbol{L}_{\mathcal{T}}^{0} \otimes \boldsymbol{L}_{\mathcal{G}}^{Q-1} \boldsymbol{y},\right. \\
\boldsymbol{L}_{\mathcal{T}}^{1} \otimes \boldsymbol{L}_{\mathcal{G}}^{0} \boldsymbol{y}, \boldsymbol{L}_{\mathcal{T}}^{1} \otimes \boldsymbol{L}_{\mathcal{G}}^{l} \boldsymbol{y}, \cdots, \boldsymbol{L}_{\mathcal{T}}^{1} \otimes \boldsymbol{L}_{\mathcal{G}}^{Q-1} \boldsymbol{y}, \\
\vdots \\
\left.\boldsymbol{L}_{\mathcal{T}}^{P-1} \otimes \boldsymbol{L}_{\mathcal{G}}^{0} \boldsymbol{y}, \boldsymbol{L}_{\mathcal{T}}^{P-1} \otimes \boldsymbol{L}_{\mathcal{G}}^{1} \boldsymbol{y}, \cdots, \boldsymbol{L}_{\mathcal{T}}^{P-1} \otimes \boldsymbol{L}_{\mathcal{G}}^{Q-1} \boldsymbol{y}\right]_{N T \times P Q},
\end{gathered}
$$
$$
\boldsymbol{C}=\left[ \begin{matrix}
	c_{0,0}&		c_{1,0}&		\cdots&		c_{P-1,0}\\
	c_{0,1}&		c_{2,1}&		\cdots&		c_{P-1,1}\\
	\vdots&		\vdots&		\ddots&		\vdots\\
	c_{0,Q-1}&		c_{1,Q-1}&		\cdots&		c_{P-1,Q-1}\\
\end{matrix} \right] ,
$$
$$
\boldsymbol{c}=\mathrm{vec}\left( \boldsymbol{C} \right) ,
$$
and rewrite (8) by using $\boldsymbol{B}$
$$
\min_{\boldsymbol{h}} \mathbb{E} \left\{ \left\| \boldsymbol{Bc}-\boldsymbol{x} \right\| _{2}^{2} \right\} . \eqno(9)
$$
\par For convenience, we define
$$
\mathcal{T} _{\lambda}^{q}=\left[ \begin{matrix}
	{\lambda _{1}^{\left( \mathcal{T} \right)}}^q&		{\lambda _{2}^{\left( \mathcal{T} \right)}}^q&		\cdots&		{\lambda _{T}^{\left( \mathcal{T} \right)}}^q\\
\end{matrix} \right] ^T,
$$
$$
\mathcal{G} _{\lambda}^{q}=\left[ \begin{matrix}
	{\lambda _{1}^{\left( \mathcal{G} \right)}}^q&		{\lambda _{2}^{\left( \mathcal{G} \right)}}^q&		\cdots&		{\lambda _{N}^{\left( \mathcal{G} \right)}}^q\\
\end{matrix} \right] ^T,
$$
$$
\boldsymbol{\Psi }_{\mathcal{T} \lambda}=\left[ \begin{matrix}
	\mathcal{T} _{\lambda}^{0}&		\mathcal{T} _{\lambda}^{1}&		\cdots&		\mathcal{T} _{\lambda}^{P-1}\\
\end{matrix} \right] _{T\times P},
$$

$$
\boldsymbol{\Psi }_{\mathcal{G} \lambda}=\left[ \begin{matrix}
	\mathcal{G} _{\lambda}^{0}&		\mathcal{G} _{\lambda}^{1}&		\cdots&		\mathcal{G} _{\lambda}^{Q-1}\\
\end{matrix} \right] _{N\times Q},
$$

$$
\begin{gathered}
\boldsymbol{\Psi}=\left[\mathcal{T}_{\lambda}^{0} \otimes \mathcal{G}_{\lambda}^{0}, \mathcal{T}_{\lambda}^{0} \otimes \mathcal{G}_{\lambda}^{2}, \cdots, \mathcal{T}_{\lambda}^{0} \otimes \mathcal{G}_{\lambda}^{Q-1},\right. \\
\mathcal{T}_{\lambda}^{1} \otimes \mathcal{G}_{\lambda}^{0}, \mathcal{T}_{\lambda}^{1} \otimes \mathcal{G}_{\lambda}^{2}, \cdots, \mathcal{T}_{\lambda}^{1} \otimes \mathcal{G}_{\lambda}^{Q-1}, \\
\vdots \\
\left.\mathcal{T}_{\lambda}^{P-1} \otimes G_{\lambda}^{0}, \mathcal{T}_{\lambda}^{P-1} \otimes G_{\lambda}^{2}, \cdots, \mathcal{T}_{\lambda}^{P-1} \otimes G_{\lambda}^{Q-1}\right]_{N T \times P Q},
\end{gathered}
$$
$$
\boldsymbol{\Psi }=\boldsymbol{\Psi }_{\mathcal{T} \lambda}\otimes \boldsymbol{\Psi }_{G\lambda},
$$
where ${\lambda _{i}^{\left( \mathcal{T} \right)}}^k$ and ${\lambda _{i}^{\left( \mathcal{G} \right)}}^k$ are the $kth$ power of the $\left( i,i \right) \mathrm{th}$ entries of $\boldsymbol{\Lambda }_{\mathcal{T}}$ and $\boldsymbol{\Lambda }_{\mathcal{G}}$, and $\boldsymbol{\Psi }_{\mathcal{T} \lambda},\boldsymbol{\Psi }_{\mathcal{G} \lambda}$ are the Vandermonde matrices.
\par The solution of (9) can be obtained by solving
$$
\mathbb{E} \left\{ \boldsymbol{B}^H\boldsymbol{B} \right\} \boldsymbol{c}=\mathbb{E} \left\{ \boldsymbol{B}^H \boldsymbol{x} \right\} . \eqno(10)
$$
Let $\boldsymbol{R}_{y,y}^{Tv}=\mathbb{E} \left\{ \boldsymbol{B}^H\mathbf{B} \right\} $, $\boldsymbol{r}_{x,y}^{Tv}=\mathbb{E} \left\{ \boldsymbol{B}^H \boldsymbol{x} \right\} $ and equation (10) is transformed as
$$
\boldsymbol{R}_{y,y}^{Tv}\boldsymbol{c}=\boldsymbol{r}_{x,y}^{Tv}. \eqno(11)
$$
\par Similar to [25], $\boldsymbol{R}_{y,y}^{Tv}$ and $\boldsymbol{r}_{x,y}^{Tv}$ are defined as autocorrelation and cross-correlation matrix on the time-vertex graph. Equation (11) is the Wiener-Hopf equation on the time-vertex graph. When $T=1$, (11) reduces to the ordinary Wiener-Hopf equation on the static graph. 
\par The entry of $\boldsymbol{R}_{y,y}^{Tv}$ at $\left( i-1 \right) Q+l\mathrm{th}$ row and $\left( j-1 \right) Q+k\mathrm{th}$ column is 
$$
\begin{aligned}
&{\left[\boldsymbol{B}^{H} \boldsymbol{B}\right]_{(i-1) Q+l,(j-1) Q+k}=\boldsymbol{y}^{H}\left(\boldsymbol{L}_{\mathcal{T}}^{i} \otimes \boldsymbol{L}_{\mathcal{G}}^{l}\right) *\left(\boldsymbol{L}_{\mathcal{T}}^{j} \otimes \boldsymbol{L}_{\mathcal{G}}^{k}\right) \boldsymbol{y}} \\
&=\boldsymbol{y}_{\mathcal{F}}^{H}\left(\boldsymbol{\Lambda}_{\mathcal{T}}^{i} \otimes \boldsymbol{\Lambda}_{\mathcal{G}}^{l}\right)*\left(\boldsymbol{\Lambda}_{\mathcal{T}}^{j} \otimes \boldsymbol{\Lambda}_{\mathcal{G}}^{k}\right) \boldsymbol{y}_{\mathcal{F}} \\
&=\boldsymbol{y}_{\mathcal{F}}^{H}\left(\left(\boldsymbol{\Lambda}_{\mathcal{T}}^{i *} \boldsymbol{\Lambda}_{\mathcal{T}}^{j}\right) \otimes\left(\boldsymbol{\Lambda}_{\mathcal{G}}^{l *} \boldsymbol{\Lambda}_{\mathcal{G}}^{k}\right)\right) \boldsymbol{y}_{\mathcal{F}} \\
&=\sum_{n=0}^{M N-1} \boldsymbol{\lambda}_{n}\left|\boldsymbol{y}_{\mathcal{F}}(n)\right|^{2}
\end{aligned}
$$
and the entry of $\boldsymbol{r}_{x,y}^{Tv}$ at $\left( i-1 \right) Q+l\mathrm{th}$ row is
$$
\begin{aligned}
&\left( \boldsymbol{B}^H \boldsymbol{x} \right) _{\left( i-1 \right) Q+l}=\boldsymbol{y}^H\left( \boldsymbol{L}_{T}^{i}\otimes \boldsymbol{L}_{\mathcal{G}}^{l} \right) ^H \boldsymbol{x}
\\
&=\boldsymbol{y}_{\mathcal{F}}^{H}\left( \boldsymbol{\Lambda} _{T}^{i}\otimes \boldsymbol{\Lambda} _{\mathcal{G}}^{l} \right) ^H \boldsymbol{x}_{\mathcal{F}}
\\
&=\sum_{n=0}^{MN-1}{\boldsymbol{\eta }_n \boldsymbol{y}_{\mathcal{F}}^{H}\left( n \right) \boldsymbol{x}_{\mathcal{F}}\left( n \right)}
\end{aligned}
$$
where $\boldsymbol{\lambda }_n=\mathrm{diag}\left( \left( \overline{\mathcal{T} _{\lambda}^{i}}\odot \mathcal{T} _{\lambda}^{j} \right) \otimes \left( \overline{\mathcal{G} _{\lambda}^{l}}\odot \mathcal{G} _{\lambda}^{k} \right) \right) $, $\boldsymbol{\eta }_n=\mathrm{diag}\left( \left( \boldsymbol{\Lambda }_{\mathcal{T}}^{i}\otimes \boldsymbol{\Lambda }_{\mathcal{G}}^{l} \right) ^H \right) $, $ \boldsymbol{y}_{\mathcal{F}}$ and $ \boldsymbol{x}_{\mathcal{F}}$ are the joint graph Fourier representation of $\boldsymbol{y}$ and $ \boldsymbol{x}$.
\par Furthermore, we rewrite the autocorrelation and cross-correlation matrix as
$$
\boldsymbol{R}_{y,y}^{Tv}=\boldsymbol{\Psi }^H\boldsymbol{Y}_{\mathcal{F}}\boldsymbol{\Psi }
$$
$$
\boldsymbol{r}_{y,y}^{Tv}=\boldsymbol{\Psi }^H\boldsymbol{Y}_{\mathcal{F}}\boldsymbol{x}_{\mathcal{F}},
$$
and the optimal coefficients $\boldsymbol{c}$ is
$$
\boldsymbol{c}=\left( \boldsymbol{\Psi }^H\boldsymbol{Y}_{\mathcal{F}}\boldsymbol{\Psi } \right) ^{-1}\boldsymbol{\Psi }^H\boldsymbol{Y}_{\boldsymbol{F}}\boldsymbol{x}_{\mathcal{F}},
$$
where $\boldsymbol{Y}_{\mathcal{F}}=\mathrm{diag}\left( \left| \boldsymbol{y}_{\mathcal{F}}\left( n \right) \right|^2,n=1,\cdots ,PQ \right) $

\section{Optimal Fractional Fourier Graph Filter on Temporal Graph Signals}

In this section, we first introduce the joint fractional Fourier transform on the time-vertex graph. Then based on the GFSO, we construct the optimal fractional Fourier time-vertex graph filter.
\noindent \subsection{The Joint Time-vertex Fractional Fourier Transform}
To capture the fractional frequency of $\boldsymbol{X}$ along both temporal and graph domains, we combine the GFRFT and discrete fractional Fourier transform to define a joint time-vertex fractional Fourier transform as
$$
\mathrm{JFRFT}\left\{ \boldsymbol{X} \right\} =\boldsymbol{U}_{\mathcal{T}}^{\left( a \right)}\boldsymbol{XU}_{\mathcal{G}}^{\left( b \right) H}. \eqno(13)
$$
Using Kronecker product, (13) can be expressed as
$$
\mathrm{JFRFT}\left\{ \boldsymbol{X} \right\} =\left( \boldsymbol{U}_{\mathcal{T}}^{\left( a \right)}\otimes \boldsymbol{U}_{\mathcal{G}}^{\left( b \right)} \right) \boldsymbol{x}=\boldsymbol{V}^{\left( a,b \right)}\boldsymbol{x}.\eqno(14)
$$
For $a=1$ and $b=1$, the JFRFT reduces to JFT in (2). For $a=0$ and $b=0$, the JFRFT reduces identity matrix.
\par \emph{Property}: JFRFT is an inversible and unitary transform.
\par \emph{Proof}: Note that $\boldsymbol{U}_{\mathcal{T}}^{\left( a \right)}$ and $\boldsymbol{U}_{\mathcal{G}}^{\left( b \right)}$ are unitary matrix. Then, we have
$$
\begin{aligned}
&\boldsymbol{V}^{\left( a,b \right)}\cdot \boldsymbol{V}^{\left( a,b \right) H}
\\
&=\left( \boldsymbol{U}_{\mathcal{T}}^{\left( a \right)}\otimes \boldsymbol{U}_{\mathcal{G}}^{\left( b \right)} \right) \left( \boldsymbol{U}_{\mathcal{T}}^{\left( a \right) H}\otimes \boldsymbol{U}_{\mathcal{G}}^{\left( b \right) H} \right) 
\\
&=\left( \boldsymbol{U}_{\mathcal{T}}^{\left( a \right)}\boldsymbol{U}_{\mathcal{T}}^{\left( a \right) H} \right) \otimes \left( \boldsymbol{U}_{\mathcal{G}}^{\left( b \right)}\boldsymbol{U}_{\mathcal{G}}^{\left( b \right) H} \right) =\boldsymbol{I}_{NT}.
\end{aligned}
$$
\subsection{The Fractional Fourier Time-vertex Graph Filter}
In this subsection, we utilize the graph fractional shift operator to construct the optimal fractional Fourier time-vertex graph filter. 
\par Similar to joint graph filter in (3), our fractional Fourier time-vertex graph filter with $\boldsymbol{L}_{\mathcal{T}}^{\left( a \right)}$ and $\boldsymbol{L}_{\mathcal{G}}^{\left( b \right)}$ is expressed as
$$
\boldsymbol{H}=h\left( \begin{array}{c}
	\boldsymbol{L}_{\mathcal{T}}^{\left( a \right)},\boldsymbol{L}_{\mathcal{G}}^{\left( b \right)}\\
\end{array} \right) =\sum_{p=0,q=0}^{P-1,Q-1}{c_{p,q}\boldsymbol{L}_{\mathcal{T}}^{\left( a \right) p}\otimes \boldsymbol{L}_{\mathcal{G}}^{\left( b \right) q}}, \eqno(15)
$$
and its joint frequency response is 
$$
h\left( \begin{array}{c}
	\boldsymbol{\Lambda }_{\mathcal{T}}^{a},\boldsymbol{\Lambda }_{\mathcal{G}}^{b}\\
\end{array} \right) =\sum_{p=0,q=0}^{P-1,Q-1}{c_{p,q}\boldsymbol{\Lambda }_{\mathcal{T}}^{ap}\otimes \boldsymbol{\Lambda }_{\mathcal{G}}^{bp}}.
$$
The optimal fractional Fourier time-vertex graph filter is expressed as
$$
\min_{\boldsymbol{H}} \mathbb{E} \left\{ \left\| \boldsymbol{Hy}-\boldsymbol{x} \right\| _{2}^{2} \right\} ,
$$
where $
\boldsymbol{H}=h\left( \begin{array}{c}
	\boldsymbol{L}_{\mathcal{T}}^{\left( a \right)},\boldsymbol{L}_{\mathcal{G}}^{\left( b \right)}\\
\end{array} \right) =\sum_{p=0,q=0}^{P-1,Q-1}{c_{p,q}\boldsymbol{L}_{\mathcal{T}}^{\left( a \right) p}\otimes \boldsymbol{L}_{\mathcal{G}}^{\left( b \right) q}}
$, $\boldsymbol{y}=\mathrm{vec}\left\{ \boldsymbol{Y} \right\} $, and $\boldsymbol{x}=\mathrm{vec}\left\{ \boldsymbol{X} \right\} $.
\par Take the conclusion in section III, and optimal coefficients of the fractional graph filter are given by
$$
\mathbf{R}_{y,y}^{FrTv}\mathbf{c}=\mathbf{r}_{x,y}^{FrTv} .\eqno(17)
$$
where $\mathbf{R}_{y,y}^{FrTv}$ and $\mathbf{r}_{x,y}^{FrTv}$ are the fractional version. When $a=1$, $b=1$, (17) is reduced to the ordinary Wiener-Hopf equation in the time-vertex graph. When $T=1$, $b=1$, (17) is reduced to the form in [25]. 
\section{The Implementation of Optimal Graph Filters on Time-vertex Graphs}
	In this section, we discuss how to implement our graph filter in sensor network data denoising. 
\par When using equation (12) to solve the optimal coefficients, we require the inverse of $\boldsymbol{\Psi }^H\boldsymbol{Y}_{\mathcal{F}}\boldsymbol{\Psi }$. $ \boldsymbol{\Psi }_{\mathcal{T} \lambda}$ and $ \boldsymbol{\Psi }_{\mathcal{G} \lambda}$ are the Vandermonde square matrices for $P=T, Q=N$ and $\boldsymbol{\Psi }^{-1}=\boldsymbol{\Psi }_{\mathcal{T} \lambda}^{-1}\otimes \boldsymbol{\Psi }_{\mathcal{G} \lambda}^{-1}$. However, $ \boldsymbol{\Psi }_{\mathcal{G} \lambda}$ is an ill-condition matrix, for a large $N$, and it is difficult to obtain its inverse matrix. Besides, when $P<T, Q<N$, $\boldsymbol{\Psi }^H\boldsymbol{Y}_{\mathcal{F}}\boldsymbol{\Psi }$ is also an ill-condition matrix. To avoid this problem, we take the method in [25] to transform the GFSO into an energy-preserving graph shift operator. 
\par For a large $T$, we need to divide $\boldsymbol{X}\in \mathbb{R} ^{N\times T}$ into $S$ groups and each length of the group is $M=T/S$ successive instants. Before using the optimal graph filter, we use the filtered graph signal by Tikhonov regularization or the recursive graph median filter to obtain the spectrum $\boldsymbol{\tilde{x}}_{\mathcal{F}}$ of estimated signal $\boldsymbol{\tilde{x}}$. We denote $\boldsymbol{\tilde{x}}$ as the preprocessing signal or the first filtered signal and our proposed graph filter as the second filtered signal. We will implement the first filtering and second filtering operation for each group to obtain the final results, respectively.
\section{NUMERICAL RESULTS}
In this section, we first show the filtering performance of different fractional orders in different datasets to prove the superiority of fractional domains over ordinary domains. Then, we present our numerical evaluations of the proposed filter and compare them with the OFrFS graph filter. The Tikhonov regularization and the recursive graph median filter are taken as the first filtering tools. All experiments are implemented on the three real-world datasets. 
\par The three real-world datasets used in our experiments are:
\par 1)	Global sea-level pressure dataset. The global sea-level pressure dataset \cite{seatemp} was published by the Joint Institute for the Study of the Atmosphere and Ocean. We select 50 nodes over a time period of 120. The selected nodes are on the world from $30^{\circ}$ north to $0^{\circ}$ and from $110^{\circ}$ east to $170^{\circ}$ east. A graph is constructed by the 5-nearest neighbors algorithm, shown in Fig.1(a). This dataset will be denoted as slp.
\par 2)	The Sea Surface Temperature dataset. The sea surface temperature dataset \cite{slp} was published by the Earth System Research Laboratory. Fifty nodes are selected over a time period of 120. The selected nodes are on the Pacific Ocean from $30^{\circ}$ south to $60^{\circ}$ south and from $170^{\circ}$ west to $90^{\circ}$ west. A graph is constructed by the 5-nearest neighbors algorithm, shown in Fig.1(b). This dataset will be denoted as sea$\_$temp. 
\par 3)	The Daily Mean PM2.5 Concentration of Texas. The Texas daily mean PM2.5 concentration dataset \cite{epa} was published by the US Environmental Protection Agency. It was collected by 47 observation sites from January 1, 2020, to December 31, 2020. Not all days have valid data, and we chose the first 120 days as our simulation data. The graph is constructed by the 5-nearest neighbors algorithm, shown in Fig. 1(c). This dataset will be denoted as PM2.5.

\begin{figure}[H]
\centering
\subfigure[]{
\includegraphics[width=4in]{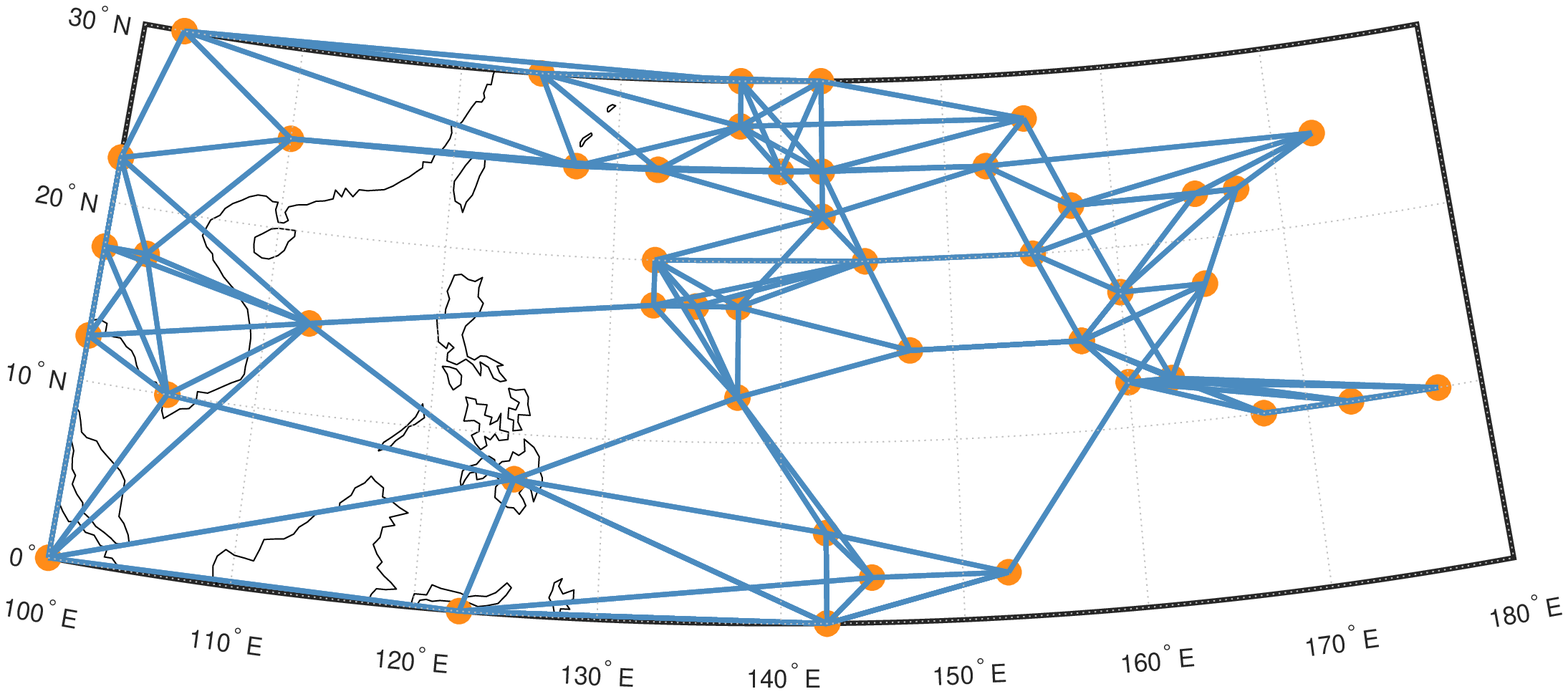}
}
\subfigure[]{
\includegraphics[width=4in]{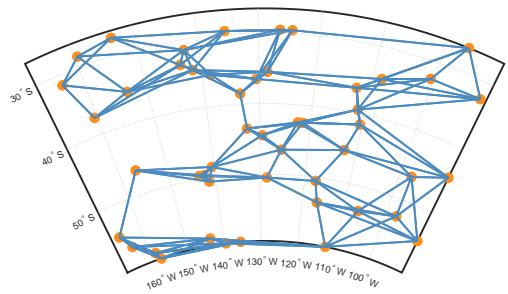}
}
\subfigure[]{
\includegraphics[width=4in]{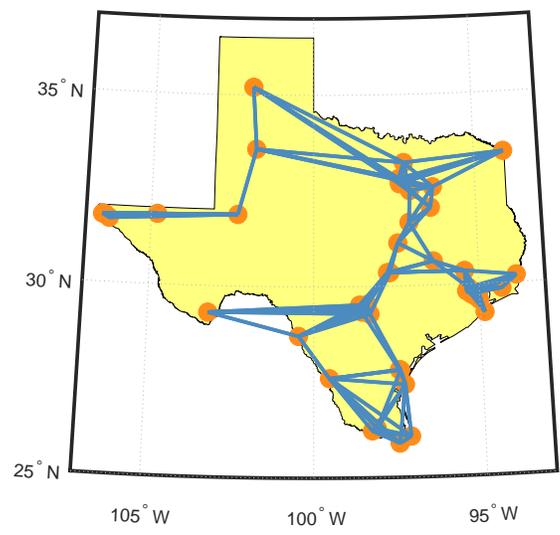}
}
\caption{Sensor networks with three datasets. (a): Global sea-level pressure dataset. (b): The Sea Surface Temperature dataset. (c): The Daily Mean PM2.5 Concentration of Texas.}
\end{figure}

\par In the experiments, we use SNR, $10\log _{10}\left( \left\| \boldsymbol{X} \right\| _F/\left\| \boldsymbol{X}-\boldsymbol{\tilde{X}} \right\|_F \right) $ or $10\log _{10}\left( \left\| \boldsymbol{X} \right\| _F/\left\| \boldsymbol{X}-\boldsymbol{Y} \right\| _F \right) $, to assess the noise strength in noisy graph signals or filtered graph signals. 

\subsection{The filtering performance with different fractional orders}
\par Fig.2 illustrates the filtering performance of our proposed filters based on the Tikhonov regularization and median graph filter in different datasets. The ‘median filter first’ represents using the recursive graph median filter as the first filtering and the ‘Tikhonov regularization first’ represents using the Tikhonov regularization as the first filtering. The black dot in each sub-image indicates the maximum SNR. The SNR of the input graph signal is -2dB. From Fig.2, we can observe that the maximum SNR is not obtained at $a=1$, $b=1$, and $a=0$, $b=0$; hence, the time-vertex optimal graph filter in the fractional domains can perform better than on the ordinary domains.

\begin{figure}[H]
\centering
\includegraphics[height=6in,width=6in]{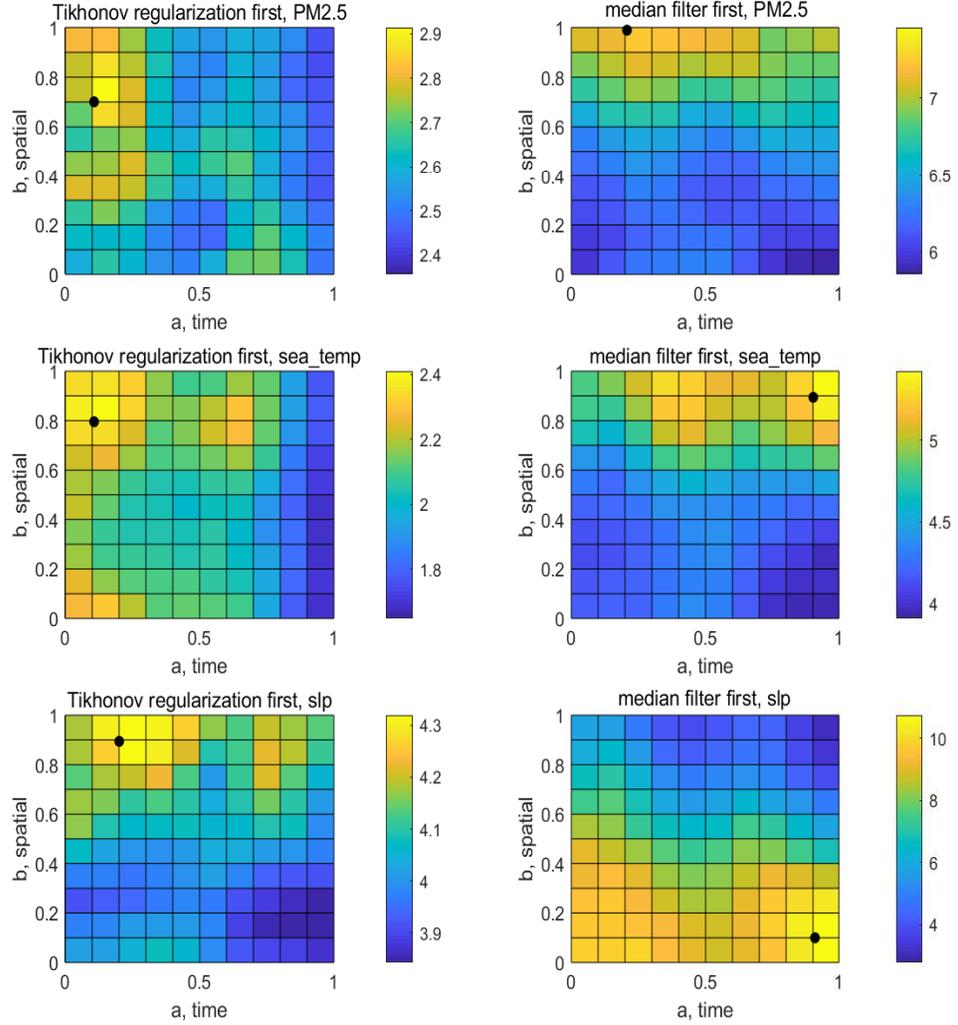}
\caption{The output SNR of our proposed graph filter versus a and b for different datasets.}
\end{figure}

\subsection{Denoising graph signals}
In this experiment, we set $P=5,T=6$, $Q=42,N=50$ in our graph filter, and $Q=42,N=50$ in the static graph optimal filter. Each SNR value is an average of 50 independent trials. The denoising performances for different graph filters are shown in Fig.3. From the observation, we can obtain that when the SNR of the input signal is less than 5dB, the result of the OFrFTv graph filter is better than that of the OFrFS graph filter. When the input SNR is higher than 5dB, the result of our filter is still better than that of the static graph filter, though the difference between them narrowed significantly.

\begin{figure}[H]
\centering
\subfigure{
\includegraphics[width=4.6in]{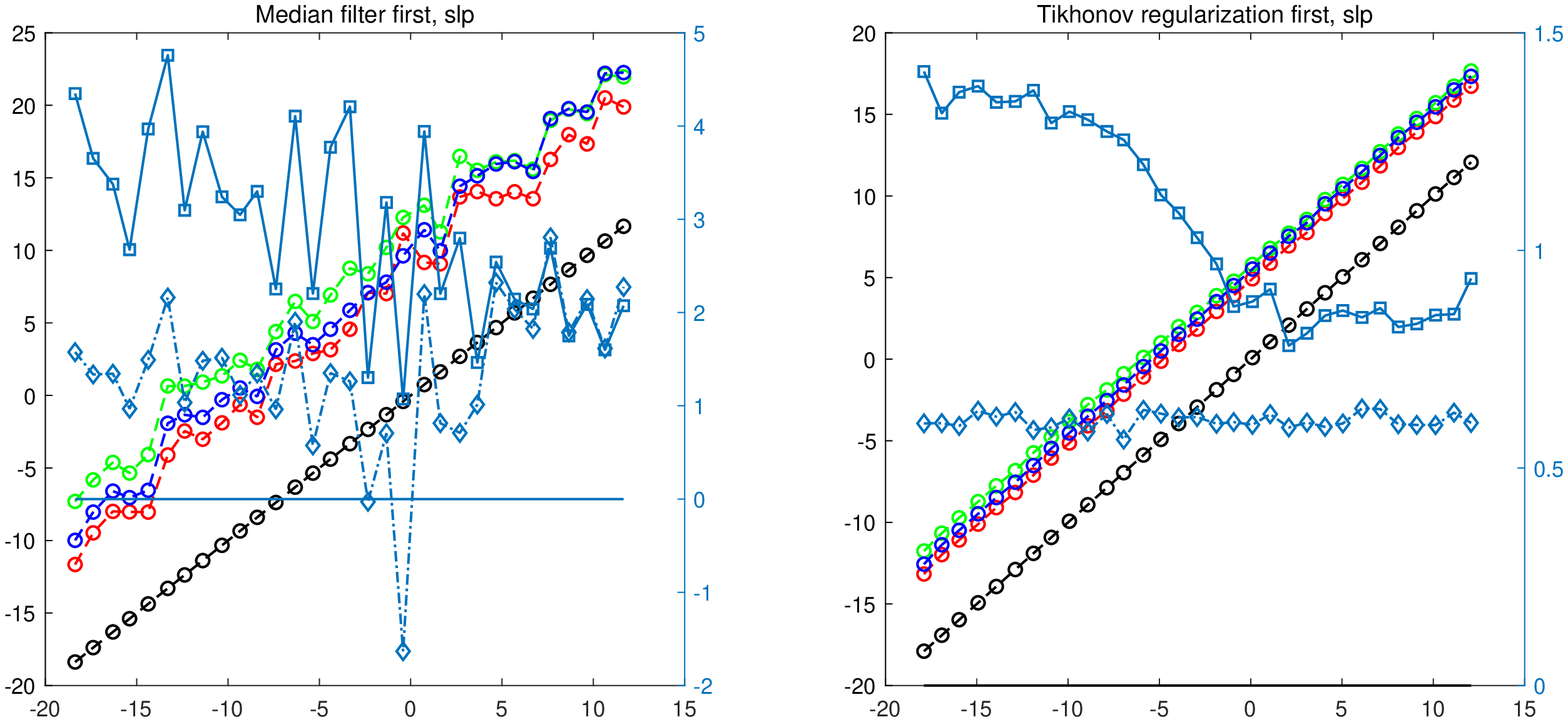}
}
\subfigure{
\includegraphics[width=4.6in]{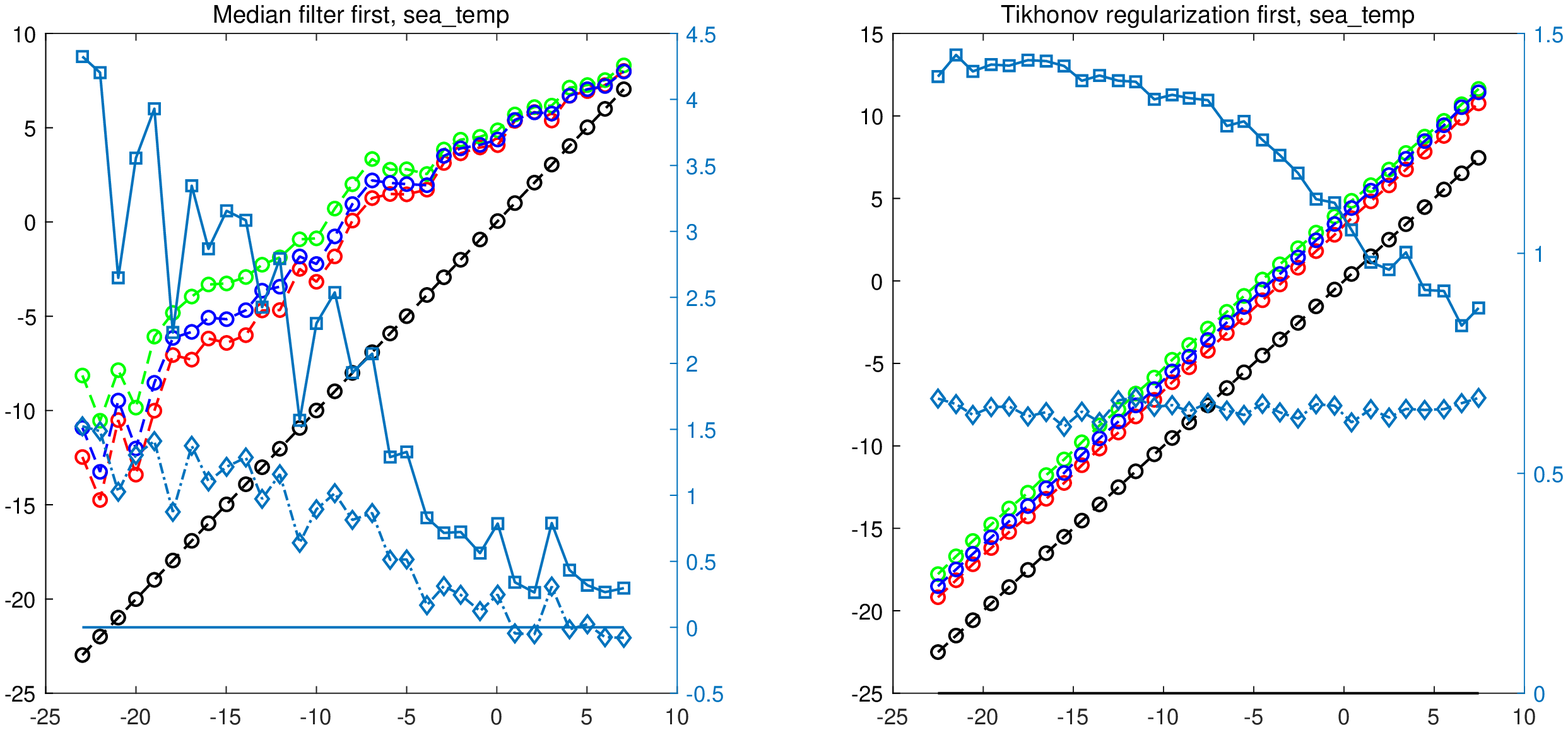}
}
\subfigure{
\includegraphics[width=4.6in]{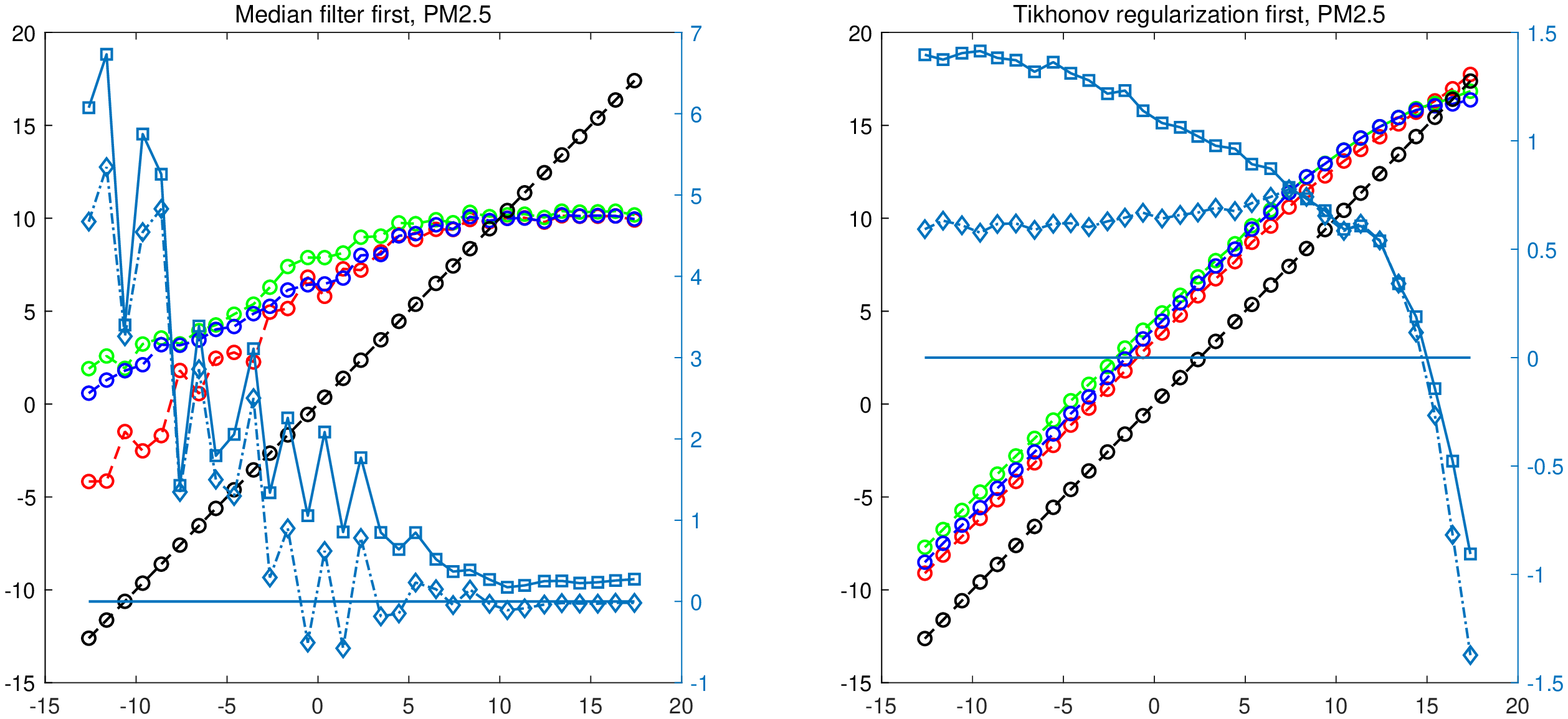}
}
\caption{Performance comparison of the proposed filter with the fractional Fourier optimal graph filter in \cite{Ref41} on different datasets. Black: the input SNR, Red: the SNR after the first filtering, Blue: the SNR after the OFrFS graph filter, Green: the SNR after our graph filter. The differences between the second filtered signal and the first filtered signal are shown in the right vertical axis. The square marker SNR is the green line subtracts the red line, and the diamond marker SNR is the blue line subtracts the red line.}
\end{figure}

\section{CONCLUSION}
\par In this paper, we first propose an optimal time-vertex graph filter design method and present the Wiener-Hopf equation in the time-vertex graph version. Then, we use the FGSO and GFRFT to extend the optimal graph filter in ordinary time/space or frequency domains to fractional domains. The designed filter was applied to denoising three time-varying real-world datasets. The results showed the superiority of fractional domains over the ordinary domains and the superiority of the time-vertex optimal filter in fractional domains over the static graph optimal filter in fractional domains.

\section*{References}
\bibliography{mybibfile}

\end{document}